\documentclass{emulateapj}
\slugcomment{draft v2}

\shorttitle{X-ray-powered Macronovae}
\shortauthors{Kisaka, Ioka \& Nakar}

\begin{document}


\title{X-ray-powered Macronovae}


\author{Shota Kisaka\altaffilmark{1}}
\email{kisaka@post.kek.jp}
\author{Kunihito Ioka\altaffilmark{1,2}}
\email{kunihito.ioka@kek.jp}
\author{Ehud Nakar\altaffilmark{3}}
\email{udini@wise.tau.ac.il}


\altaffiltext{1}{Theory Center, Institute of Particle and Nuclear Studies, KEK, Tsukuba 305-0801, Japan}
\altaffiltext{2}{Department of Particle and Nuclear Physics, SOKENDAI (The Graduate University for Advanced Studies), Tsukuba 305-0801, Japan}
\altaffiltext{3}{The Raymond and Beverly Sackler School of Physics and Astronomy, Tel Aviv University, Tel Aviv 69978, Israel}


\begin{abstract}
A macronova (or kilonova) was observed as an infrared excess
several days after short gamma-ray burst, GRB 130603B.
Although the $r$-process radioactivity is widely discussed as an energy source,
it requires huge mass of ejecta from a neutron star (NS) binary merger.
We propose a new model that the X-ray excess gives rise to the simultaneously observed infrared excess via thermal re-emission 
and explore what constraints this would place on the mass and velocity of the ejecta.
This X-ray-powered model explains both the X-ray and infrared excesses with a single energy source by the central engine like a black hole,
and allows for broader parameter region,
in particular smaller ejecta mass $\sim10^{-3}-10^{-2}M_{\odot}$ with iron mixed 
as suggested by general relativistic simulations for typical NS-NS mergers,
than the previous models.
We also discuss the other macronova candidates in GRB 060614 and GRB 080503,
and implications for the search of electromagnetic counterparts to gravitational waves.
\end{abstract}





\section{INTRODUCTION}
\label{sec:introduction}

What is the energy source of a macronova\footnote{
Following the usage of ``supernova'', which does not specify the energy source,
we use the term ``macronova'' 
for a transient associated with the neutron star (NS) binary mergers,
whatever the energy source is.
} (or kilonova)?
Macronovae are considered as the emission from the ejecta with mass $\sim10^{-4} - 10^{-1}M_{\odot}$ 
and the velocity $\sim0.1-0.3c$, which are accompanied by the mergers of a NS binary
\footnote{We use a term ``a binary NS" for a NS-NS binary and ``a NS binary" for a NS-NS or black hole-NS (BH-NS) binary.} 
\citep[e.g., ][]{LP98, K05, Hot+13, Kyutoku:2013wxa, Kyu+15}.
A NS binary merger is one of the most promising sources for the direct detection of the gravitational wave (GW).
The identification of the electromagnetic counterpart to a GW source 
\citep{MB12,R15}
would significantly reduce the huge localization error of the GW detectors 
\citep[$\sim10-100$ deg$^2$; e.g., ][]{Ber+15, Ess+15} 
such as Advanced LIGO \citep{LIGO15}, Advanced VIRGO \citep{VIRGO15} and KAGRA \citep{KAGRA13}. 
The strategy of follow-up observations of GW sources should be improved
by clarifying the main energy source of macronovae, which determines the behavior of the light curve. 

Nuclear heating due to the decay of the $r$-process elements has been widely discussed 
as a heating source of macronovae \citep[e.g., ][]{LP98, Met+10, KBB13, TH13, Ros+14, Tanaka+14, Wan+14, LR15}.
Since merger ejecta are neutron-rich, 
$r$-process elements could be synthesized,
and the NS binary mergers could be the origin of
$r$-process elements \citep[e.g., ][]{LS74}, 
although iron-rich ejecta are also possible in some cases via the shock and/or neutrino heating \citep[e.g., ][]{MPQ09}.
The $r$-process elements also affect the opacity of the ejecta \citep{KBB13, TH13},
making the long timescale and low temperature of macronovae.

The activities of the central engine are also proposed as
alternative energy sources of the macronovae \citep{KIT15, KIN15}.
After a NS binary merger, either a BH or a NS is formed as a remnant.
The central remnant releases energy through
such as the relativistic jet \citep[e.g., ][]{KI15}, disk wind \citep[e.g., ][]{PZ06, KKS12, KKSSW14, Kiu+15} 
and NS wind \citep[e.g., ][]{Dai+06, MQT08, Row+13, GOWR13, GOW14, Lu+15}, 
which may be observed as prompt, extended and plateau emissions in short gamma-ray bursts (GRBs) 
\citep[e.g., ][]{N07, B14}.
These outflows can heat the ejecta, which may shine as the macronovae.
Similar energy injection is considered in the core-collapse supernovae \citep[e.g.,][]{KB10}.
The energy injection by a highly magnetized NS (magnetar) could also produce 
a brighter transient than the observed macronovae 
\citep[e.g., ][]{YZG13, F+13, MP14, WDY15, Gao+15}, although short GRBs
do not show any radio signatures of magnetars
\citep[e.g., ][]{NP11, PNR13, MB14, TKI14}.

Recently, a macronova candidate is detected as the infrared excess in GRB 130603B \citep{Tan+13, BFC13}.
From observations, the luminosity, timescale and temperature are $\sim10^{41}$ erg s$^{-1}$, $\sim7$ days and $\lesssim4\times10^3$ K, respectively.
In the $r$-process model, relatively large ejecta mass for the merger of a NS-NS binary, $\gtrsim0.03M_{\odot}$, is required to reproduce the observed infrared excess \citep[e.g., ][]{BFC13, PKR14, Gro+14}. 
This is important for the equation of state of high-density matter and the progenitor of short GRBs (BH-NS merger or NS-NS merger) \citep[e.g., ][]{Hot+13b}.
Too large ejecta mass in the $r$-process model may imply
the engine-powered model 
\citep{KIT15, KIN15}.
Still the previous models using 
the activities of the extended and plateau emissions also require the large ejecta mass $\gtrsim0.02M_{\odot}$ \citep{KIT15, KIN15}.
In addition, the previous models also require ejecta with high opacity $\sim10$ cm$^2$ g$^{-1}$, which is comparable to the $r$-process case,
to reproduce the observed long timescale and low temperature.

GRB 130603B also shows the long-lasting mysterious X-ray component, which is significantly in excess of the extrapolated power laws based on the optical afterglow \citep{Fong+14}.
The detected luminosity is $\sim10^{42}$ erg s$^{-1}$ at the observed time $\sim6\times10^5$ s, which is longer than the timescale of the plateau emission \citep{Row+13, GOWR13, GOW14, Lu+15}, and 
could originate from
the central engine such as the accretion disk emission due to the fallback material \citep[e.g., ][]{R07, RB09, Kyu+15}.
Since the central engine is surrounded by the ejecta, we expect the interaction between the emitted X-ray and the surrounding ejecta.
Then the ejecta heated by the irradiation of X-ray may emit infrared photons and reproduce the observed infrared excess according to the ejecta properties.
This is without need for any additional energy source such as the radioactive decay of $r$-process elements. 
Such an X-ray-powered model has a significant advantage that the model uses a single energy source to explain two mysterious signals - X-ray and infrared - which are observed at the same time and with similar luminosities.
This is in contrast to the $r$-process and previous engine-powered models which
require different unrelated sources for the infrared and X-ray excesses.

In this paper, we consider the X-ray-powered macronovae to explain the observed macronova candidate, GRB 130603B.
In Section \ref{sec:model} we describe our model and constraints on the ejecta properties. 
In Section \ref{sec:results},  we present the results for the macronova in GRB 130603B.
The other macronova candidates in GRB 060614 and GRB 080503
are discussed in Section \ref{sec:others}.
We present the discussion and summary in Section \ref{sec:discussion}.

\section{X-RAY-POWERED MODEL}
\label{sec:model}

 \begin{figure}
  \begin{center}
   \includegraphics[width=70mm, angle=270]{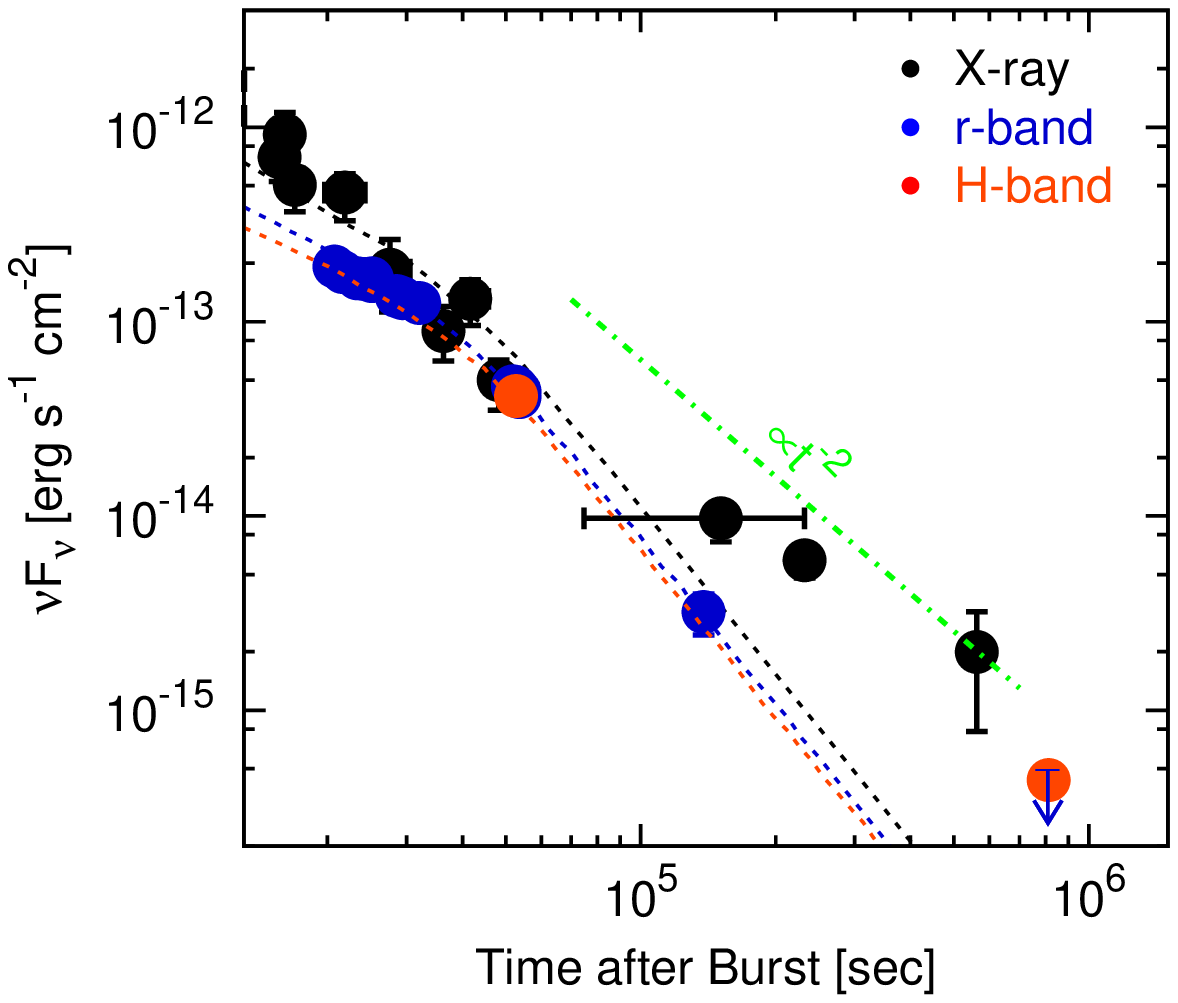}
   \caption{Observed light curves of GRB 130603B. 
Black, blue and red denote the X-ray (1 keV), optical (r-band) and infrared (H-band), respectively.
Data and lines of the afterglow model are taken from \citet{Fong+14} \citep[some of them are originated from][]{Tan+13, BFC13, Cuc+13, de+14}. We use the optical and infrared fluxes corrected by the extinction of the host galaxy in \citet{Fong+14}. The green dot-dashed line shows $L\propto t^{-2}$, on which the X-rays give the same contribution to the macronova emission. After the jet break time $t\gtrsim4\times10^4$ s, 
the X-ray and infrared fluxes are in comparable excesses
of the extrapolated power laws based on the optical emission. }
   \label{figure:lightcurve}
  \end{center}
 \end{figure}

Figure \ref{figure:lightcurve} shows the observed X-ray (1 keV; black), optical (r-band; blue) and infrared (H-band; red) light curves of GRB 130603B \citep{Tan+13, BFC13, Cuc+13, de+14, Fong+14}. 
The optical light curve shows the jet break at $\sim4\times10^4$ s \citep{Fong+14}, which is supposed to be achromatic.
However, X-ray and infrared fluxes are significantly in excess of the extrapolated power laws after $\sim2\times10^5$ s.
Remarkably, the X-ray luminosity is similar or 
slightly larger than the infrared excess
known as the macronova emission
at the time $\sim10^6$ s. 

The similar luminosities suggest the same origin 
for the X-ray and infrared excesses:
the infrared macronova could be reproduced
if a part of the X-ray excess
is converted to the infrared emission,
whatever the X-ray excess is.
This X-ray-powered model 
is naturally expected because
the NS binary mergers would eject matter of mass $M_{\rm ej}\sim10^{-3}-10^{-1}M_{\odot}$, as shown by general relativistic simulations
\citep[e.g., ][]{Hot+13, Kyutoku:2013wxa, Kyu+15, Kaw+15},
and provide a screen to absorb the X-ray and re-emit the infrared emission.
This model explains the observed infrared macronova without the nuclear heating due to the $\beta$-decay and fission of $r$-process elements \citep[e.g., ][]{LP98}.

Note that 
the activities of the extended emission (timescale $\sim10^2$ s) 
and plateau emission (timescale $\sim10^4$ s) may also 
contribute energy to the macronova
\citep{KIT15, KIN15}.
In these cases, the energy is injected with shorter timescale than that of the macronovae ($\sim10^6$ s) and remains trapped with adiabatic cooling ($E_{\rm int}\propto t^{-1}$) before being released at $\sim10^6$ s. 
As a result, 
the required luminosity for explaining the macronova
scales as $L\sim E_{\rm int}/t\propto t^{-2}$.
In Figure \ref{figure:lightcurve}, we plot $L\propto t^{-2}$ by the green dot-dashed line.
Because the X-ray excess component follows $L\propto t^{-1.88}$ \citep{Fong+14},
the later energy injection is more effective for the macronova emission.

\begin{figure}
  \begin{center}
   \includegraphics[width=75mm]{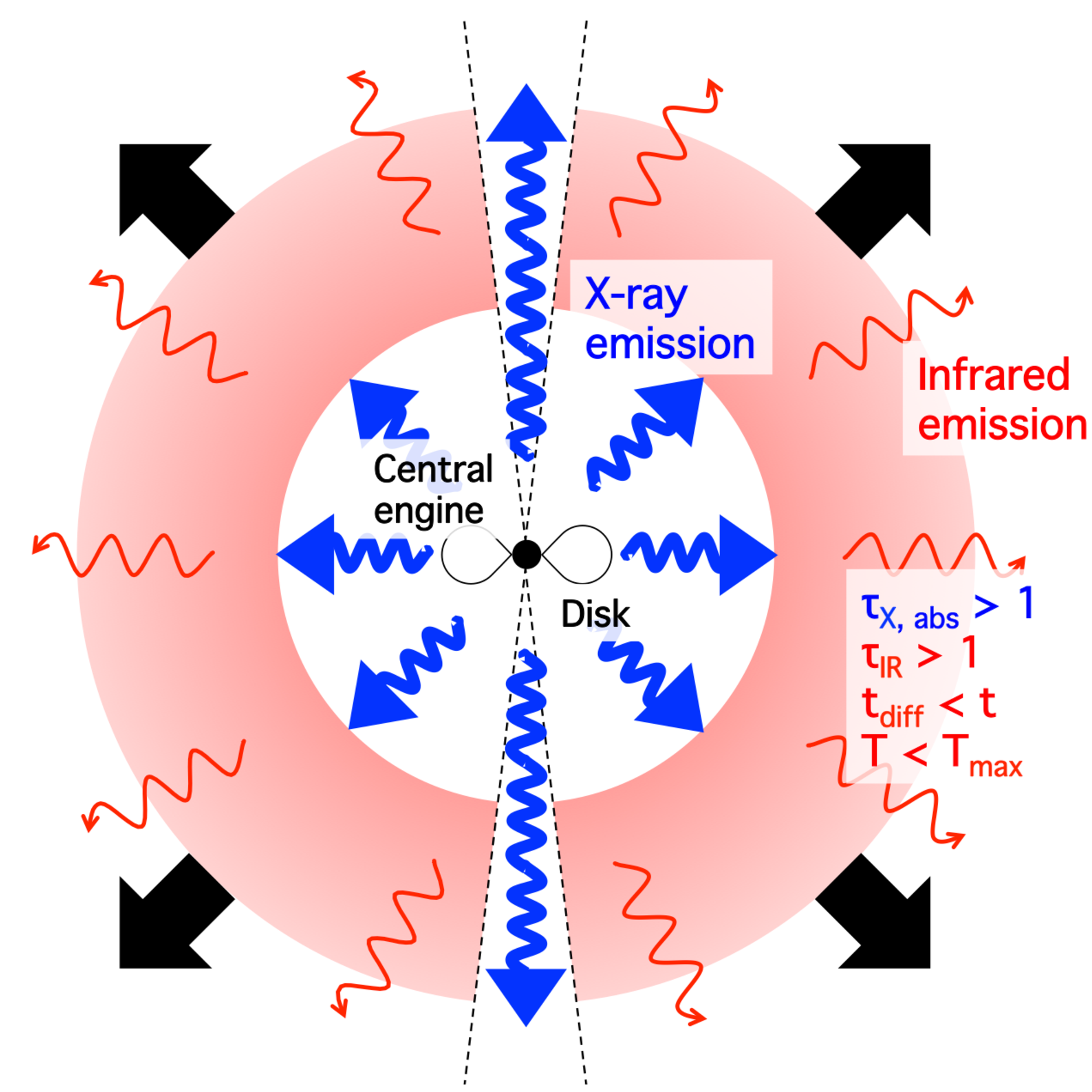}
   \caption{Schematic picture of the X-ray-powered model.}
   \label{figure:schematic}
  \end{center}
 \end{figure}

For the X-ray-powered model to produce the X-ray and infrared excesses detected by observers simultaneously, we consider the following physical setups
in Figure \ref{figure:schematic}. 
First, X-rays are generated near the central source and are emitted 
in nearly isotropic direction. Second, the ejecta lie at radius larger than the X-ray source and cover a fraction of
solid angle. Third, the line of sight to the X-ray source is clean for the observers who detect the GRB emission. 

Although our model does not depend on the specific mechanism of X-ray emission,
we assume that the X-ray excess at $\sim1-10$ days in GRB 130603B originates from the activity near the central engine, such as the accretion disk with super-Eddington accretion rate \citep[e.g., ][]{R07, RB09, FM13, FKMQ15, Fer+15, Kyu+15} like ultra-luminous X-ray sources.
At the early stage of the merger, a relativistic jet is launched from the central engine to penetrate the ejecta \citep{Nag+14, Mur+14}.
Because the jet makes a hole in the ejecta, the observers toward the jet axis can directly see inside the ejecta (see Figure \ref{figure:schematic}). 

Let us first consider nearly isotropic ejecta with a constant velocity $v_{\rm ej}$ based on the results of the numerical simulations \citep[e.g., ][]{Hot+13}.
The anisotropy of the ejecta will be discussed in Section \ref{sec:anisotropy}.
The radius of the ejecta is 
\begin{eqnarray}\label{ejecta_radius}
R_{\rm ej}\sim v_{\rm ej}t,
\end{eqnarray}
where $t$ is the time after the merger.
Since the velocity structure becomes homologous and 
the ejecta spreads to the radial direction,
the typical mass density of the ejecta is given by 
\begin{eqnarray}\label{ejecta_mass_density}
\rho_{\rm ej}\sim\frac{3M_{\rm ej}}{4\pi R_{\rm ej}^3}.
\end{eqnarray}
For the composition of the ejecta, we consider both the iron-rich ejecta and the heavy $r$-process ejecta.
Since the tidally ejected matter is neutron rich,
the heavy $r$-process elements may be synthesized \citep[e.g., ][]{LS74}. 
On the other hand, the iron-rich ejecta may be dominant
if the shocks and/or neutrino irradiation make the electron fraction high
\citep[e.g., ][]{MPQ09, Sek+15, Mar+15, Ric+15}. 

We consider four conditions to reproduce the infrared excess in our X-ray-powered model. 
First, the X-ray photons should be absorbed by the ejecta, i.e.,
the optical depth for the X-ray absorption $\tau_{\rm X, abs}$
should be larger than unity,
\begin{eqnarray}\label{absorption_condition}
\tau_{\rm X, abs}> 1.
\end{eqnarray}
Second, the absorbed X-ray photons should be thermalized in the ejecta to produce infrared photons.
Since the opacities for the iron and $r$-process elements are decreasing function of wavelength \citep{KBB13, TH13}, 
we require the optical depth for infrared photons $\tau_{\rm IR}$ to be larger than unity,
\begin{eqnarray}\label{thermalization_condition}
\tau_{\rm IR}>1.
\end{eqnarray}
Third, the infrared photons should escape from the ejecta to be detected
by the observers.
Since the optical depth $\tau_{\rm IR}$ satisfies condition (\ref{thermalization_condition}), 
we consider random walk for the propagation of infrared photons in the ejecta. 
The infrared photons can escape from the ejecta if the diffusion timescale $t_{\rm diff}$ is smaller than the dynamical timescale $t$, 
\begin{eqnarray}\label{escape_condition}
t_{\rm diff}<t.
\end{eqnarray}
In the case that the central engine is active longer than the photon diffusion timescale, 
the light curve of the thermal emission follows the evolution of the central engine, 
i.e., the peak time is not necessarily when $t\sim t_{\rm diff}$ in the X-ray-powered model. 
This is the reason why the X-ray-powered model requires smaller amount of ejecta than the previous models (see section 3).
Finally, the temperature of the ejecta $T$ should be 
lower than the observed upper limit $T_{\max}\sim4000$ K \citep{Tan+13, BFC13},
\begin{eqnarray}\label{temperature_condition}
T<T_{\max}\sim 4000\,{\rm K},
\end{eqnarray}
which is obtained from the detected infrared flux and the optical upper limit (blue arrow in Figure \ref{figure:lightcurve}).
In following Sections \ref{sec:absorption} - \ref{sec:temperature}, we consider these four conditions to give constraints on the ejecta properties.
We also consider the luminosity ratio of X-ray and infrared excesses in Section \ref{sec:anisotropy}. 

\subsection{X-ray Absorption}
\label{sec:absorption}

In this subsection, we consider condition (\ref{absorption_condition}) 
for the X-ray photons to be absorbed in the ejecta.
The absorption optical depth is given by
\begin{eqnarray}\label{absorption_optical_depth}
\tau_{\rm X, abs}\sim\kappa_{\rm bf}\frac{M_{\rm ej}}{4 \pi R_{\rm ej}^2},
\end{eqnarray}
where $\kappa_{\rm bf}$ is the bound-free opacity.
The opacity $\kappa_{\rm bf}$ depends on the ionization state of the ejecta \citep{MP14},
\begin{eqnarray}\label{bound-free_opacity}
\kappa_{\rm bf}\sim\frac{f_{\rm n}}{\bar{A}m_{\rm p}}\sigma_{\rm bf},
\end{eqnarray}
where $\bar{A}$ is the average mass number of the composed elements of the ejecta, $m_{\rm p}$ is the proton mass, $\sigma_{\rm bf}$ is the photoionization cross section and $f_{\rm n}$ is the neutral fraction of the ejecta.
The ionization state is determined by the balance of the photoionization due to the X-ray emission from the central engine and the recombination. 
The neutral fraction $f_{\rm n}$ is described by the ratio of the absorption rate of the ionizing photons ${\mathcal R}_{\rm ion}$ and the recombination rate ${\mathcal R}_{\rm rec}$ as
\begin{eqnarray}\label{neutral_fraction}
f_{\rm n}\sim\left(1+\frac{{\mathcal R}_{\rm ion}}{{\mathcal R}_{\rm rec}}\right)^{-1}.
\end{eqnarray}
The recombination rate is approximately described by 
\begin{eqnarray}\label{recombination_rate}
{\mathcal R}_{\rm rec}\sim n_{\rm e}\alpha_{\rm rec}, 
\end{eqnarray}
where $n_{\rm e}$ is the number density of the electrons and $\alpha_{\rm rec}$ is the recombination coefficient.
The ionization rate is
\begin{eqnarray}\label{ionization_rate}
{\mathcal R}_{\rm ion}\sim n_{\rm ph}\sigma_{\rm bf}c, 
\end{eqnarray}
where the number density of photons whose energy is larger than the ionization threshold energy is, 
\begin{eqnarray}\label{photon_number_density}
n_{\rm ph}\sim\frac{\epsilon L_{\rm X}}{4\pi h\nu_{\rm i} R_{\rm ej}^2c},
\end{eqnarray}
where $h\nu_{\rm i}$ is the ionization threshold energy and $\epsilon$ is 
the fraction of the luminosity above the ionization energy.

In the case of iron-rich ejecta, the number density of the electrons in the fully-ionized state is $n_{\rm e}\sim26\rho_{\rm ej}/56m_{\rm p}$, 
the recombination coefficient for the hydrogen-like iron is $\alpha_{\rm rec}\sim6.3\times10^{-10}(T/10^4{\rm K})^{-0.8}$ cm$^3$ s$^{-1}$ \citep{OF06}, 
the ionization threshold energy of the innermost electronic state is $h\nu_{\rm i}\sim9.277$ keV, 
and the photo-ionization cross section at $\nu=\nu_{\rm i}$ is $\sigma_{\rm bf}\sim1.2\times10^{-20}$ cm$^2$ \citep{OF06}. 
Using these values and the irradiated X-ray luminosity $\epsilon L_{\rm X}\sim10^{41}$ erg s$^{-1}$
\footnote{We use the X-ray luminosity at $0.3-10$ keV and regard the excess component as non-thermal emission 
because the photon index is $\sim-2$ from observational data at $\sim10^5$ s (http://www.swift.ac.uk/index.php) when the excess component dominates the afterglow emission.
Since the number of photons is smaller in higher energy in the non-thermal emission, 
we consider that the the photons with higher energy than 9.2 keV do not significantly change the ionization fraction.}, 
the ratio ${\mathcal R}_{\rm ion}/{\mathcal R}_{\rm rec}$ is much smaller than unity for the reasonable parameter ranges of the merger ejecta, 
$v_{\rm ej}/c\gtrsim10^{-2}$, $M_{\rm ej}\gtrsim10^{-3}M_{\odot}$ and $T\lesssim4000$ K, at the time $t\sim7$ days.
Then, we use $f_{\rm n}=1$. 
Substituting Equations (\ref{ejecta_radius}), (\ref{ejecta_mass_density}), (\ref{absorption_optical_depth}) and (\ref{bound-free_opacity}) with $f_{\rm n}=1$ into condition (\ref{absorption_condition}), we obtain the constraint on the ejecta velocity
\begin{eqnarray}\label{absorption_condition_for_velocity}
\frac{v_{\rm ej}}{c}&\lesssim&0.79\left(\frac{\bar{A}}{56}\right)^{-1/2}\left(\frac{t}{7\,{\rm day}}\right)^{-1}\left(\frac{M_{\rm ej}}{10^{-2}\,M_{\odot}}\right)^{1/2}.
\end{eqnarray}
As long as the neutral fraction $f_{\rm n}=1$ is a good approximation, inequality (\ref{absorption_condition_for_velocity}) does not depend on the temperature.
Note that since the photo-ionization cross section for electrons bounded at outer shell 
is larger than that for the hydrogen-like iron \citep{OF06}, lower energy ionizing photons ($\lesssim10$ keV) are also absorbed by the ejecta as long as condition (\ref{absorption_condition_for_velocity}) is satisfied. 

For the $r$-process ejecta, we assume that most X-ray photons are absorbed by the ejecta within $\sim10$ days. 

If the X-ray emission is beamed or a limited region of the ejecta is irradiated, 
the total irradiated luminosity to the ejecta is smaller than that estimated from the observed X-ray luminosity. 
Note that the ratio ${\cal R}_{\rm ion}/{\cal R}_{\rm rec} \propto t^{-3.88}/t^{-3}
\propto t^{-0.88}$
is decreasing, so that 
the neutral fraction remains unity, $f_{\rm n}=1$.

\subsection{Thermalization}
\label{sec:thermalization}

We consider condition (\ref{thermalization_condition}) for the thermalization in the ejecta.
The optical depth for optical-infrared photons is given by
\begin{eqnarray}\label{infrared_optical_depth}
\tau_{\rm IR}\sim\kappa_{\rm IR}\frac{M_{\rm ej}}{4 \pi R_{\rm ej}^2}.
\end{eqnarray}
The dominant opacity at the infrared wavelength is bound-bound opacity. 
In the iron ejecta, the opacity for the thermal photons is $\kappa_{\rm IR}\sim0.1$ cm$^2$ g$^{-1}$ \citep{KBB13, TH13}.
For simplicity, we neglect the dependence of the opacity $\kappa_{\rm IR}$
on the temperature.
From condition (\ref{thermalization_condition}) and Equations (\ref{ejecta_radius}), (\ref{ejecta_mass_density}) and (\ref{infrared_optical_depth}), the upper limit on the ejecta velocity is  
\begin{eqnarray}\label{thermalization_condition_for_velocity}
\frac{v_{\rm ej}}{c}&\lesssim&2.2\times10^{-2}\left(\frac{\kappa_{\rm IR}}{0.1\,{\rm cm}^2{\rm g}^{-1}}\right)^{1/2}\left(\frac{t}{7\,{\rm day}}\right)^{-1}\nonumber \\
& &\times\left(\frac{M_{\rm ej}}{10^{-2}\,M_{\odot}}\right)^{1/2}.
\end{eqnarray}
For the $r$-process ejecta, 
\citep[$\kappa_{\rm IR}\sim10$ cm$^2$ g$^{-1}$; ][]{KBB13, TH13}, the upper limit on the ejecta velocity becomes large. 

\subsection{Diffusion}
\label{sec:diffusion}

Thermal photons can escape from the ejecta if condition (\ref{escape_condition}) is satisfied.
The diffusion timescale $t_{\rm diff}$ for the propagation distance $R_{\rm ej}$ is 
\begin{eqnarray}\label{diffusion_timescale}
t_{\rm diff}\sim\tau_{\rm IR}\frac{R_{\rm ej}}{c}.
\end{eqnarray}
Then, condition (\ref{escape_condition}) and Equations (\ref{ejecta_radius}), (\ref{ejecta_mass_density}), (\ref{infrared_optical_depth}) and (\ref{diffusion_timescale}) give the lower limit on the velocity, 
\begin{eqnarray}\label{escape_condition_for_velocity}
\frac{v_{\rm ej}}{c}&\gtrsim&4.8\times10^{-4}\left(\frac{\kappa_{\rm IR}}{0.1\,{\rm cm}^2{\rm g}^{-1}}\right)\left(\frac{t}{7\,{\rm day}}\right)^{-2}\nonumber \\
& &\times\left(\frac{M_{\rm ej}}{10^{-2}\,M_{\odot}}\right).
\end{eqnarray}

\subsection{Temperature}
\label{sec:temperature}

Since the upper limit on the temperature was obtained from the observations at optical and infrared bands \citep{Tan+13, BFC13}, the temperature of the ejecta has to satisfy condition (\ref{temperature_condition}) to reproduce the observations 
\footnote{Although the lower limit on the temperature is also obtained by using the upper limit on the flux density in 6.7 GHz presented by \citet{Fong+14}, the derived constraint on the ejecta velocity is trivial ($v_{\rm ej}<c$).}.
The temperature of the ejecta is determined by the relation 
\begin{eqnarray}\label{internal_energy}
4\pi R_{\rm ej}^2\sigma T^4\sim L_{\rm IR},
\end{eqnarray}
where $\sigma$ is Stefan-Boltzmann constant.
Then, using condition (\ref{temperature_condition}) and Equation (\ref{internal_energy}), the lower limit on the ejecta velocity is 
\begin{eqnarray}\label{temperature_condition_for_velocity}
\frac{v_{\rm ej}}{c}&\gtrsim&4.1\times10^{-2}\left(\frac{T_{\max}}{4000\,{\rm K}}\right)^{-2}\left(\frac{L_{\rm IR}}{10^{41}\,{\rm erg~s}^{-1}}\right)^{1/2}\nonumber \\
& &\times\left(\frac{t}{7\,{\rm day}}\right)^{-1}.
\end{eqnarray}
Since the temperature is determined by the internal energy density due to the irradiation, the lower limit on the velocity (\ref{temperature_condition_for_velocity}) does not depend on the ejecta mass. 

\subsection{X-ray and infrared Luminosities}
\label{sec:anisotropy}

In Figure \ref{figure:lightcurve}, the X-ray luminosity is 
approximately an order of magnitude
larger than the infrared luminosity at $t\sim10^6$ s.
The difference between the X-ray and infrared luminosities may come from 
the collimated X-ray emission due to the geometry of the super-Eddington accretion disk \citep[e.g., ][]{TO15}. 
Note that the ejecta velocity is non-relativistic, so that the reprocessed infrared emission is isotropic.

The anisotropy of the ejecta could also 
reduce the infrared luminosity compared to the X-ray luminosity.
In the BH-NS mergers, the dynamical ejecta are highly anisotropic \citep[e.g., ][]{Kyutoku:2013wxa, Kyu+15, Kaw+15}. 
If the ejecta cover $\sim$10\% of the solid angle of the X-ray emission, the luminosity of the reprocessed emission from the ejecta is roughly expected to be $L_{\rm IR}\sim0.1L_{\rm X}$. 

The anisotropy of the ejecta composition could also contribute to the difference between X-ray and infrared luminosities.
Numerical simulations including the neutrino transport showed that the electron fraction around the orbital axis is relatively higher than that near the equatorial plane \citep{Sek+15, Mar+15, Ric+15}.
Then, iron-rich ejecta may only concentrate on the orbital axis. 
In the other region, $r$-process elements may be synthesized and hide the reprocessed photons with the high opacity. 
This effect effectively reduces the emission area of the ejecta so that the thermal luminosity becomes lower than the X-ray luminosity. 

Conversely, the infrared luminosity may be larger than the observed X-ray luminosity because of the limited energy band of the observations. 
If a significant energy is emitted as unobserved soft X-rays, 
the infrared reemission has larger luminosity than 
the observed X-rays.

\section{RESULTS}
\label{sec:results}

 \begin{figure*}
  \begin{center}
   \includegraphics[width=170mm]{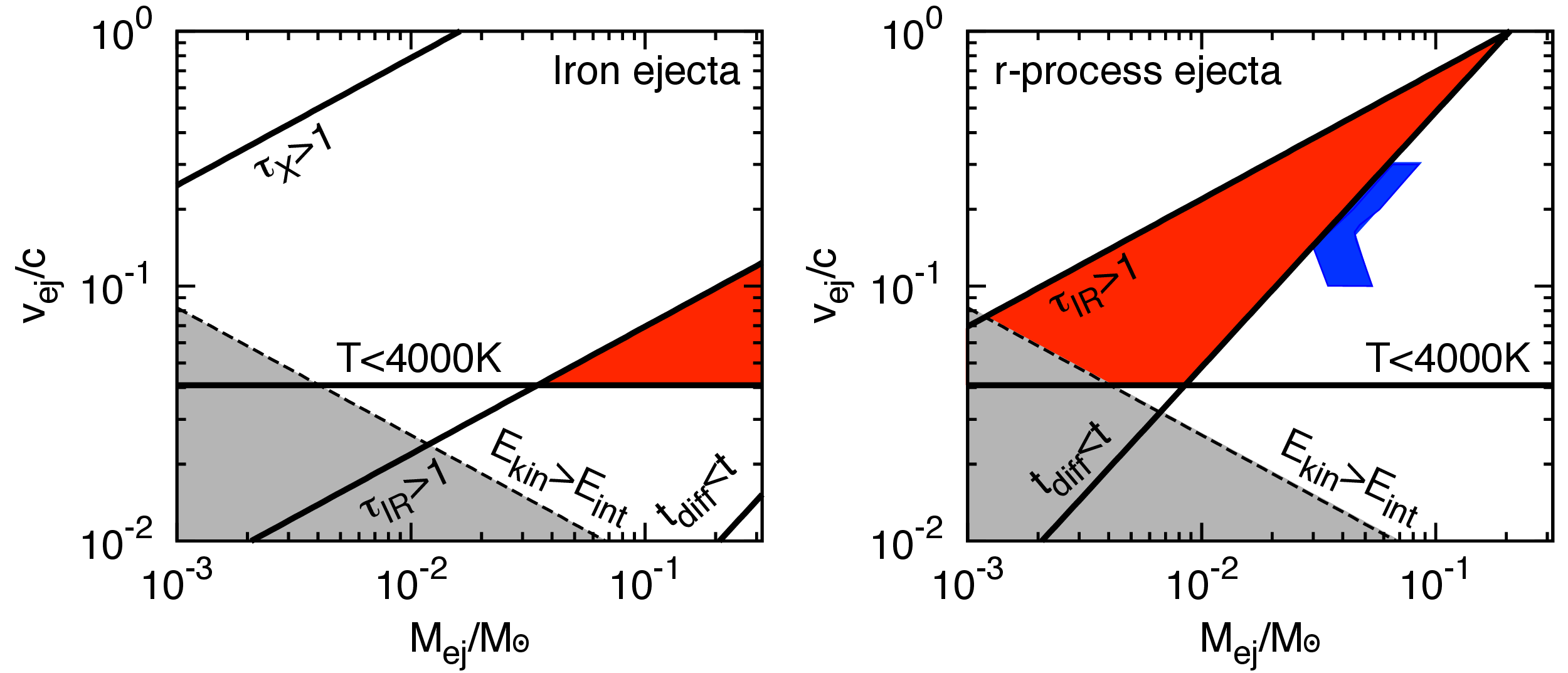}
   \caption{Allowed parameter space to reproduce the infrared excess by reprocessing the X-ray excess emission in GRB 130603B. 
Left panel shows the iron ejecta case. 
A reasonable range of the parameters, the ejecta mass $M_{\rm ej}\gtrsim0.04M_{\odot}$ and velocity $v_{\rm ej}/c \gtrsim 0.04 $,
satisfies our model criteria. 
The condition ${\cal R}_{\rm rec}>{\cal R}_{\rm ion}$ is satisfied in the displayed area with $h\nu_{\rm i}=10$ keV and $\epsilon L_{\rm X}\sim10^{41}$ erg s$^{-1}$, so that $f_{\rm n}=1$ is a good approximation (Equation (\ref{neutral_fraction})).
The kinetic energy is larger than the internal energy above the grey region
[the boundary (the dashed line) is described by Equation (\ref{ejecta_velocity})],
which is the physical situation.
Right panel shows the $r$-process ejecta case. 
The ejecta mass $M_{\rm ej}\sim10^{-3}-10^{-2}M_{\odot}$ and velocity $v_{\rm ej}/c\sim0.04-0.2$
satisfy our model criteria.
For comparison, we also plot the parameter region
of the $r$-process heating model in \citet{BFC13}
as a blue region. 
}
   \label{figure:parameter}
  \end{center}
 \end{figure*}

In the left panel of Figure \ref{figure:parameter}, we show the allowed parameter space (red region) for the iron ejecta in the X-ray-powered model. 
We use $f_{\rm n}=1$, $\bar{A}=56$, $T=4\times10^3$ K,  $t=7$ days, $\kappa_{\rm IR}=0.1$ cm$^2$ g$^{-1}$ and $L_{\rm IR}=10^{41}$ erg s$^{-1}$. 
The luminosity, temperature and timescale are consistent with the observed values in GRB 130603B \citep{Tan+13, BFC13}. 
Conditions (\ref{absorption_condition_for_velocity}), (\ref{thermalization_condition_for_velocity}), (\ref{escape_condition_for_velocity}) and (\ref{temperature_condition_for_velocity}) are plotted as solid lines.
This panel shows that even if the ejecta are composed of iron, the X-ray-powered model can explain the observed infrared excess 
by the reasonable range of the parameters ($M_{\rm ej}\gtrsim4\times10^{-2}M_{\odot}$ and $v_{\rm ej}/c\sim0.04-0.1$). 

The right panel of Figure \ref{figure:parameter} shows the case that the ejecta is dominated by the $r$-process elements. 
The X-ray absorption condition requires such as the recombination rate and photoionization cross section of the $r$-process elements. 
Here, we assume that $r$-process elements are partially ionized (not all f-shell electrons are ionized)
and most X-ray photons are absorbed by the ejecta within $\sim10$ days. 
Then, the bound-bound opacity is significant for the optical-infrared photons. 
We use $\kappa_{\rm IR}\sim10$ cm$^2$ g$^{-1}$ \citep{KBB13, TH13}. 
Other parameters are the same as in the iron ejecta case in the left panel of Figure \ref{figure:parameter}. 
The right panel of Figure \ref{figure:parameter} shows that even if the ejecta mass is $M_{\rm ej}\sim10^{-3}-10^{-2} M_{\odot}$, the observed infrared excess is reproduced in our X-ray-powered model with the ejecta velocity $v_{\rm ej}/c\sim0.04-0.2$. 

In the right panel of Figure \ref{figure:parameter}, we also plot the allowed region for the $r$-process nuclear heating model in \citet{BFC13} (blue region).
This panel shows that the allowed region of the parameter space (red region) is much larger than that of the $r$-process heating model\footnote{
See \citet{KIT15} for more precise calculations of
the allowed parameters in the $r$-process heating model.
}. 
Since the $r$-process heating model gives the peak luminosity at the time $t\sim t_{\rm diff}$ \citep[e.g., ][]{LP98}, the allowed region of parameters resides near the line derived from condition (\ref{escape_condition}). 
The X-ray-powered model (red region) has lower ejecta mass than the $r$-process model, so that the contribution of the $r$-process heating is negligible in the X-ray-powered model. 

In the both panels of Figure \ref{figure:parameter}, we plot dashed lines where the kinetic energy of the ejecta $E_{\rm kin}$ equals to the internal energy of the ejecta injected by the irradiation $E_{\rm int}\sim L_{\rm IR}t$ at the time $t=7$ days. 
If the kinetic energy of the ejecta before the irradiation is smaller than the internal energy provided by the irradiation (the grey region in Figure \ref{figure:parameter}), the ejecta would be accelerated and the both energies reach the equipartition, $E_{\rm kin}=E_{\rm int}$. 
This means that the ejecta velocity after the irradiation is determined by the internal energy deposited by the irradiation $E_{\rm int}\sim L_{\rm IR}t$ and the ejecta mass $M_{\rm ej}$, 
\begin{eqnarray}\label{ejecta_velocity}
v_{\rm ej}\sim\sqrt{\frac{2 L_{\rm IR}t}{M_{\rm ej}}},
\end{eqnarray}
as long as the velocity is non-relativistic.
We should use this velocity for deriving four conditions 
in the case $E_{\rm kin}<E_{\rm int}$ \footnote{In the range shown in Figure \ref{figure:parameter}, most of the regions $E_{\rm kin}=E_{\rm int}$ (dashed line) are excluded by the temperature condition (\ref{temperature_condition_for_velocity}).}.
In the $r$-process ejecta, the allowed region with the ejecta mass $M_{\rm ej}\lesssim10^{-3}M_{\odot}$ satisfies $E_{\rm kin}<E_{\rm int}$.
If we consider the ejecta with the mass $M_{\rm ej}<10^{-3}M_{\odot}$ and the initial velocity $v_{\rm ej}$ which satisfies conditions $\tau_{\rm IR}>1$ (\ref{thermalization_condition_for_velocity}) and $T<T_{\max}$ (\ref{temperature_condition_for_velocity}), the ejecta velocity after the irradiation, Equation (\ref{ejecta_velocity}), exceeds the upper limit from condition $\tau_{\rm IR}>1$ (\ref{thermalization_condition_for_velocity}).
Therefore, the lower limit on the ejecta mass is $M_{\rm ej}\sim10^{-3}M_{\odot}$ in the case of the $r$-process ejecta.
On the other hand, for the mass range $10^{-3}M_{\odot}\lesssim M_{\rm ej}\lesssim4\times10^{-3}M_{\odot}$, 
even if the initial ejecta velocity $v_{\rm ej}$ is smaller than the limit from $T<T_{\max}$,
the irradiation accelerates the velocity to the allowed region.
Note that $E_{\rm int}\sim L_{\rm IR}t$ is the lower limit on the injected internal energy because in the early stage the larger energy may be injected, 
which gives negligible emission at the time $\sim10$ days due to the adiabatic cooling \citep{KIT15, KIN15}. 

\section{On the other macronova candidates}
\label{sec:others}

\subsection{Application to GRB 060614}

The model of X-ray-powered macronovae could also applicable to the macronova candidate following GRB 060614. 
As is the case in GRB 130603B, GRB 060614 has the long-lasting ($\sim10^6$ s) X-ray components whose timescale and flux are comparable to those of infrared excess components \citep{Man+07, Yang+15, Jin+15}. 
The luminosity, temperature and timescale are $\sim10^{41}$ erg s$^{-1}$, $\sim2200-3400$ K and $\sim12$ day in the source rest frame, respectively 
\citep{Yang+15, Jin+15}. 
Because of relatively long timescale ($\sim$1.7 times longer than that of the macronova in GRB 130603B), large ejecta mass $M_{\rm ej}\sim0.1M_{\odot}$ is required in the $r$-process heating model \citep{Yang+15, Jin+15}. 
Such ejecta mass may be too large for
NS-NS mergers as suggested by the simulations
\citep[e.g., ][]{Hot+13}. 
Thus, \citet{Yang+15} and \citet{Jin+15} suggested a BH-NS merger 
\citep{Kyutoku:2013wxa,Kyu+15} for the origin of GRB 060614.
On the other hand, if we apply our X-ray-powered model to the observed excesses in GRB 060614, 
the required minimum mass is
$\sim 0.1 M_{\odot}$ for the iron ejecta and $\sim2\times10^{-3}M_{\odot}$ for the $r$-process ejecta, where
the thermalization and photon escape conditions (\ref{thermalization_condition_for_velocity}) and (\ref{escape_condition_for_velocity}) 
give the scaling $M_{\rm ej}\propto t^2$ compared to GRB 130603B.
The required ejecta velocity $v_{\rm ej}/c\sim0.03-0.08$\footnote{Although the velocity is slightly small compared with the escape velocity of the neutron stars ($\sim0.2c$), the uncertainties of the temperature which come from the assumed afterglow model may reduce the difference.} is the same as GRB 130603B (from condition (\ref{temperature_condition_for_velocity})). 
Therefore, a NS-NS merger can explain the observed macronova in GRB 060614 
more naturally in our X-ray-powered model.

\subsection{Application to GRB 080503}

The X-ray-powered model may also explain the optical rebrightening at $\sim1-5$ day after the GRB 080503.
The long-lasting X-ray ($\sim1-2$ day) has a similar flux to the optical one \citep{Per+09}. 
Unfortunately, the redshift of GRB 080503 is only limited to $z<4$ \citep{Per+09}.
From the observations \citep{Per+09}, the optical flux and temperature are $\sim10^{-15}$ erg s$^{-1}$ and $\sim5\times10^3 - 10^4$ K, respectively, which are roughly constant between $\sim1 - 5$ days in the observer frame. 
To explain this event based on the X-ray-powered model, the thermalization condition (\ref{thermalization_condition_for_velocity}) at $\sim5$ days and the escape condition (\ref{escape_condition_for_velocity}) at $\sim1$ day have to be satisfied.
We find that the small mass $M_{\rm ej}\sim10^{-4}M_{\odot}$ and the velocity $v_{\rm ej}/c\sim0.01-0.04$ are required in the case of the $r$-process dominated ejecta.
The redshift is also restricted to $z<0.1$, otherwise
the kinetic energy of the ejecta is smaller than the observed radiated energy, which is unphysical.
On the other hand, the iron ejecta with the mass $M_{\rm ej}\sim3\times10^{-3}-10^{-2}M_{\odot}$ and the velocity $v_{\rm ej}/c\sim0.01-0.04$ can also explain the observed optical rebrightening in GRB 080503 within the redshift range $z \lesssim 0.5$. 
The allowed parameter region for GRB 080503 with the iron ejecta slightly overlaps with that for GRB 130603B with the $r$-process ejecta, but not with the iron ejecta.
If GRB 080503 and GRB 130603B have the same ejecta mass and velocity, 
our results imply that the ejecta composition is not isotropic. 
Namely, some parts of the ejecta are dominated by iron while other parts are dominated by $r$-process elements. 
This picture is supported by the numerical simulations \citep[e.g., ][]{Sek+15, Mar+15, Ric+15}.

\section{SUMMARY AND DISCUSSION}
\label{sec:discussion}

We propose the X-ray-powered macronovae as an alternative interpretation of the observed infrared excess in the short GRB 130603B. 
Our model has important advantages over the $r$-process heating model \citep[e.g., ][]{LP98} and the previous engine-powered models \citep{KIT15, KIN15}. 
First, the X-ray-powered model explains both the X-ray and infrared excess components with similar fluxes using a single energy source. 
On the other hand, the previous models must consider different unrelated sources to explain the excess components. 
Second, the X-ray-powered model allows for a much broader parameter region of the ejecta properties than that of the $r$-process model (in the right panel of Figure \ref{figure:parameter}). 
In particular, the X-ray-powered model explains the macronova with smaller ejecta mass.  
The $r$-process model must satisfy all the conditions which the X-ray-powered model satisfies, in addition to having the correct amount of $r$-process material to generate the observed infrared luminosity by radioactive decay,
as well as to having the correct diffusion time at the observed peak time $t \sim t_{\rm diff}$.
Then, the allowed parameter range of the ejecta mass has to be much narrow in the $r$-process model. 
In the previous engine-powered models \citep{KIT15, KIN15}, the energy injection occurs at $10^2-10^4$ s, earlier than the macronova emission. 
To confine the injected internal energy in the ejecta up to the macronova timescale $\sim10^6$ s, the ejecta satisfy the condition $t<t_{\rm diff}$ until $\sim10^6$ s. 
This condition always requires the larger ejecta mass than that of the X-ray-powered model.

We also apply the X-ray-powered model to the other macronova candidates, GRB 060614 and GRB 080503. 
The $r$-process ejecta with the mass $M_{\rm ej}\sim2\times10^{-3}-3\times10^{-2}M_{\odot}$ and the velocity $v_{\rm ej}/c\sim0.03-0.08$ can explain the optical-infrared excess in GRB 060614. 
As oppose to the $r$-process heating model \citep{Yang+15, Jin+15}, 
the X-ray-powered model allows relatively small ejecta mass expected for the NS-NS mergers \citep[e.g., ][]{Hot+13}. 
In GRB 080503,
the allowed parameter range does not overlap with that in GRB 130603B
expect for the case that GRB 080503 has the iron ejecta while
GRB 130603B has the $r$-process ejecta.
This may imply ejecta that are in part made of iron and in part made of $r$-process elements, as suggested by numerical simulations that find anisotropic electron fraction in the ejecta of NS-NS mergers.

In the case that the temperature in the ejecta becomes low ($\lesssim2000$ K), dust may form in the ejecta depending on the ejecta composition \citep{TNI14}. 
The dust enhances the opacity of the ejecta.
Then, the thermalization occurs for lower mass density of the ejecta 
than Figure \ref{figure:parameter} (condition (\ref{thermalization_condition_for_velocity})).
On the other hand, photons can escape in the ejecta only with lower mass density (condition (\ref{escape_condition_for_velocity})).
As a result, the allowed region moves toward lower ejecta mass
than Figure \ref{figure:parameter}. 

\citet{MP14} suggested that ejecta is fully ionized and X-ray photons freely escape after $\sim$ 1 day for their magnetar model with the X-ray luminosity $L_{\rm X}\sim10^{45}$ erg s$^{-1}$. 
In our model, the irradiated X-ray luminosity is much smaller than their magnetar model ($L_{\rm X}\sim10^{45}$ erg s$^{-1}$), so that the ejecta are not fully ionized. 

As the X-ray excess component, we consider the activity near the central engine, such as the accretion disk. 
Using the fallback model with the temporal evolution of the mass accretion rate $\dot{M}\propto t^{-5/3}$ at $t>t_0$ and the radiative efficiency $\sim0.1$, 
the luminosity from the disk is $\sim10^{41}$ erg s$^{-1}~(M_{\rm f}/0.017M_{\odot})(t_0/0.1~{\rm s})^{2/3}(t/7~{\rm day})^{-5/3}$, 
where $M_{\rm f}$ is the total fallback mass \citep[e.g., ][]{KI15}. 
Note that the ratio of the fallback mass $M_{\rm f}$ to the ejecta mass $M_{\rm ej}$ is $\sim1-10$ in our model, 
which is consistent with the numerical simulations \citep{Hot+13c, Hot+13, Nag+14, Kyu+15}. 
If the central engine activity comes from the magnetar, 
the luminosity is $\sim10^{41}$ erg s$^{-1}~(B_{\rm d}/2\times10^{16}~{\rm G})^2(P/0.45~{\rm s})^{-4}$ 
where $B_{\rm d}$ is the surface dipole magnetic field and the $P$ is the rotation period of the NS at 7 days \citep[e.g., ][]{U92}.
Since the X-ray excess component roughly follows the temporal index $\sim-2$, the period $P$ has been slowed down at 7 days. 

There are important implications of the X-ray-powered macronovae for the search of the electromagnetic counterparts to GW sources.
If we can first observe the X-ray emission originated from the central engine
\citep{Kan+12, KIN15}, 
the thermal emission from the ejecta could be estimated from the X-ray emission as the luminosity 
\begin{eqnarray}
L\sim L_{\rm X},
\end{eqnarray}
and the temperature 
\begin{eqnarray}
T&\sim&3000~\left(\frac{v_{\rm ej}}{0.07c}\right)^{-1/2}\left(\frac{L}{10^{41}~{\rm erg~s}^{-1}}\right)^{1/4} \nonumber \\
& &\times\left(\frac{t}{7~{\rm day}}\right)^{-1/2}{\rm K},
\end{eqnarray}
as long as the ejecta satisfy the conditions (\ref{absorption_condition})$-$(\ref{escape_condition}).
Even if the ejecta initially prevent the detection of X-ray photons from the central engine, the ejecta become optically thin to the X-ray tens of days after the merger.
The luminosity is about $\sim10^{40}-10^{41}$ erg s$^{-1}$, which could be detectable by {\it Swift}/XRT with integration time $10^4$ s within 100 Mpc \citep{Mor+07}.

If the X-ray from the central engine is significantly faint, 
other heating sources would become dominant such as the $r$-process 
and the shorter timescale activities of the central engine (extended and plateau emissions). 
Thus, the X-ray observations are important to distinguish the main heating source of the macronovae.

\acknowledgments
We are grateful to the anonymous referee for useful suggestions. 
We would also like to thank
K. Asano, K. Hotokezaka, T. Piran, R. Sari 
for fruitful discussions. 
This work is supported by 
SOKENDAI (The Graduate University for Advanced Studies),
KAKENHI 24103006 (S.K., K.I.), 24000004, 26247042, 26287051 (K.I.). E.N. was partially supported by an ERC starting grant (GRB/SN), an ISF grant (1277/13), an ISA grant and an I-CORE Program (1829/12).

\end{document}